\documentstyle[dina4,12pt,epsf]{article}
\begin{document}

\setlength{\parindent}{0.0cm}
\epsfxsize=12cm

\begin{center}
  {\LARGE
    {\bf
      {\baselineskip1.0cm
        An Algorithm for \\
        Dynamic Load Balancing \\
        of Synchronous Monte Carlo Simulations \\
        on Multiprocessor Systems \\
      }
    }
  }

\vspace{3cm}

  \renewcommand{\thefootnote}{\fnsymbol{footnote}}
  Peter Altevogt\footnote[1]{altevogt@dhdibm1.bitnet},
  \renewcommand{\thefootnote}{\arabic{footnote}}
  Andreas Linke \\
  Institute for Supercomputing and Applied Mathematics (ISAM) \\
  Heidelberg Scientific Center \\
  IBM Deutschland Informationssysteme GmbH \\
  Vangerowstr. 18 \\
  69115 Heidelberg \\
  Tel.: 06221--59--4471 \\
  Germany

\end{center}

\vspace{4cm}

\begin{abstract}
  We describe an algorithm for dynamic load balancing
  of geometrically parallelized synchronous Monte Carlo
  simulations of physical models. This algorithm is designed
  for a (heterogeneous) multiprocessor system of the MIMD
  type with distributed memory. The algorithm is based on a
  dynamic partitioning of the domain of the algorithm, taking
  into account the actual processor resources of the various
  processors of the multiprocessor system. \\\\

\end{abstract}
\vfill
\hfill IBM preprint 75.93.08, hep-lat/9310021

\newpage

\section{Introduction}

During the last years,
Monte Carlo simulations of scientific problems have turned out
to be of outstanding importance \cite{Creutz1, Bi0, Bi1, Bi2, VLSC1}.
It is now a general belief within the community of
computational scientists that
multiprocessor systems of the MIMD\footnote{Multiple
Instruction stream, Multiple Data stream.
For an excellent overview on the various
models of computation like SISD, SIMD, etc., see \cite{Akl1}.} type with
distributed memory are the
most appropriate computer systems to provide the computational resources needed
to solve the most demanding problems e.g.\ in High Energy Physics, Material
Sciences or Meteorology. \\

Implementing a synchronous parallel algorithm on a heterogeneous MIMD system
with distributed memory (e.g.\ on
a cluster of different workstations), a load balancing between the processors
of the system (taking into account the actual resources being available on
each node) turns out to be crucial, because the processor with the
least resources determines the speed of the complete algorithm.  \\

The heterogenity of the MIMD system may not only result
because of heterogeneous hardware resources,
but also due to a {\em heterogeneous use}
of homogeneous hardware resources (e.g.\ on a workstation cluster, there
may exist several serial tasks running on some of the workstations of the
cluster for some time in addition to the parallel application; this
results in a temporary
heterogenity of the cluster, even if the workstations of the cluster are
identical). This kind of heterogenity can in general only
be detected during the runtime of the parallel algorithm. \\

Therefore, the usual approach of
{\em geometric parallelization} \cite{He2, Alt1} to parallelize algorithms
by a {\em static}
decomposition of a domain into subdomains and associating each
subdomain with a processor of the multiprocessor system is
not appropriate for a heterogeneous multiprocessor system. Instead, on a
heterogeneous system this {\em geometric parallelization} should be
done {\em dynamic}. \\

In the sequel,
we consider geometrically parallelized Monte Carlo simulations consisting of
update algorithms
(e.g. Metropolis or heatbath algorithms) defined on e.g. spins
at the sites of a lattice (e.g. in Ising models), matrices defined on the links
of the lattice (e.g. in Lattice Gauge Theories), etc.. In general, a
synchronization of the parallel processes takes place after each sweep through
the lattice (each iteration). \\

For this class of simulations,
we will introduce an algorithm for dynamic load balancing.
Implementing and testing
the algorithm for the two--dimensional Ising model, it will be shown,
that this algorithm may drastically improve
the performance of the Monte Carlo simulation on a heterogeneous multiprocessor
system. \\

The paper consist of basically three parts. In the first part we will introduce
a simple model for analyzing
the performance of synchronous Monte Carlo simulations
on multiprocessor systems with distributed memory, in the second part we will
present our algorithm
for the dynamic load balancing and finally we will present
our numerical results. \\

\section{The Performance Model}

We consider a heterogeneous\footnote{In the sense of the introduction.}
multiprocessor system consisting of $n$ processors. For a
parallelized Monte Carlo
simulation we measure\footnote{Using e.g. a library routine provided
by the operating system.} at times $t_{MC}$ the times
$\Delta t_i^{t_{MC}}$ the simulation has taken for
a fixed number of sweeps on each of the processors\footnote{Here the times
$\Delta t_i$ denote
``wall clock times'' measured in seconds and $t_{MC}$ denotes
the ``internal'' time of the simulation, measured in numbers of Monte Carlo
sweeps.}. The parallelization is done by
{\em geometric parallelization},
associating a sublattice $i$ with a characteristic
scale $L_i^{t_{MC}}$
(e.g. a characteristic side of the sublattice, its volume, etc..)
with each of the processors. These scales are choosen such that

\begin{eqnarray}
L_i^{t_{MC}} = c_i^{t_{MC}} L \label{LIDEF}
\end{eqnarray}

holds. Here $L$ denotes
the scale of the complete lattice and the $c_i^{t_{MC}}$ are real numbers with
$0 < c_i^{t_{MC}} < 1$
and $\sum_i c_i^{t_{MC}} = 1$\footnote{The initial choice of the
parameters ${L_i^{t_{MC}=0}}$ respectively of the ${c_i^{t_{MC}=0}}$ is quite
arbitrary, one could
choose e.g. ${c_i^{t_{MC}=0}} = \frac{1}{n}$ for all $i$.}.
Using these parameters,
we can calculate quantities $P_i^{t_{MC}}$, characterizing the computing
resources of processor $i$ at time $t_{MC}$:

\begin{eqnarray}
  P_i^{t_{MC}} := \frac{c_i^{t_{MC}}}{\Delta t_i^{t_{MC}}}  \label{PIDEF}
\end{eqnarray}

Assuming, that the resources of the processors vary slowly compared
with the time the simulation takes for one sweep

\begin{eqnarray}
  P_i^{t_{MC}+1} \sim P_i^{t_{MC}}, \label{PSIMP}
\end{eqnarray}

we set

\begin{eqnarray}
  P_i = P_i^{t_{MC}} = const . \label{PICONST}
\end{eqnarray}

Now we reinterpret
formula (\ref{PIDEF}): For {\em fixed} $P_i$ we want to calculate
a set of $\{c_i^{t_{MC}+1}\}$, such that the time for the next
sweep ({\em excluding} the time spent for communication)\footnote{From now
on throughout this section we always consider the times $\Delta t$,
$\Delta t_i$, etc. and the coefficients $\{c_i\}$ to be taken at $t_{MC}+1$.
For the sake of clarity we therefore drop this
index throughout this section.}

\begin{eqnarray}
  \Delta t(\{c_i\}) := \max_{i} \Delta t_i(c_i) \label{NSWEEPT}
  \qquad {\rm with} \qquad  \Delta t_i(c_i) := \frac{c_i}{P_i}
\end{eqnarray}
%
%

for $i = 1,...,n$ has a minimal value

\begin{eqnarray}
  \Delta t_{min} := \min_{\{c_i\}}
                                 \Delta t(\{c_i\}). \label{TMINDEF}
\end{eqnarray}

A necessary condition for this solution is obviously, that all
$\Delta t_i$ must be equal\footnote{Let us assume that e.g.
$\Delta t_1>\Delta t_2$. Then
we could easily make $\Delta t_1$ smaller by a redefinition
of $c_1$ and $c_2$ with $c_1 + c_2 = const$.}. Remembering the
normalization condition on the constants $c_i$, we arrive at

\begin{eqnarray}
  c_i = \frac{P_i}{\sum_i P_i}, \label{CISOL}
\end{eqnarray}

with $i=1,...,n$. Using (\ref{NSWEEPT}) this results in

\begin{eqnarray}
  \Delta t_{min} = \frac{1}{\sum_i P_i}. \label{TMINSOL}
\end{eqnarray}

For a homogenous system (all $P_i$ equal) (\ref{CISOL}) would give

\begin{eqnarray}
 c_i = \frac{1}{n} \label{CISOLHOM}
\end{eqnarray}

For a heterogenous system this choice of $\{c_i\}$ results in
(using (\ref{NSWEEPT}))

\begin{eqnarray}
  \Delta t(\{c_i=\frac{1}{n}\}) = \frac{1}{n} \frac{1}{\min_i P_i}.
  \label{TNINSOL}
\end{eqnarray}

As the {\em homogenity} $H$ of the multiprocessor system we define
therefore the ratio of $\Delta t_{min}$ with (\ref{TNINSOL}):

\begin{eqnarray}
  H = n\frac{\min_{i} P_{i}}{\sum_{i} P_{i}} \label{DEFHOM}
\end{eqnarray}

with $0\le H \le 1$. The speedup $S$ that can be obtained by the
dynamic resizing of the sublattices is the inverse of $H$:

\begin{eqnarray}
  S = \frac{1}{H}. \label{SPEEDUPDEF}
\end{eqnarray}

Rewriting (\ref{TMINSOL}) in terms of $H$, we arrive at

\begin{eqnarray}
  \Delta t_{min} = \frac{1}{n} \frac{H}{\min_i P_i}, \label{TMINSOLH}
\end{eqnarray}

\begin{figure}[h]
  \centering
  \leavevmode
  \epsffile{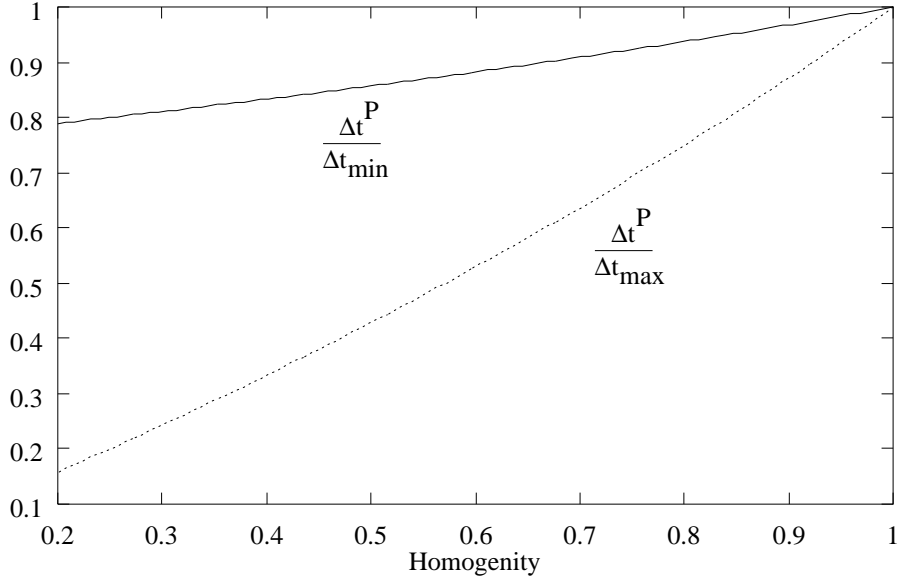}
  \caption{Performance predicted by our model for 4 processors with
           and without load balancing. \label{theory}}
\end{figure}

For later comparison with our experimental data,
let us look at the special case
$P := P_1 = ... = P_{n-1} > P_n =: P_{min}$. In this case we have for
$\Delta t_{min}$:

\begin{eqnarray}
  \Delta t_{min} = \Delta t^P \frac{n - H}{n - 1}, \label{TMINSOLHS}
\end{eqnarray}

with
$\Delta t^P = \frac{1}{nP}$\footnote{Using $P_{min}=\frac{(n-1)H}{n-H}P$.}.
Without load balancing, the time for the total simulation
will be determined by the time spent on processor $n$:
\begin{eqnarray}
  \Delta t_{max} & = & \frac{1}{nP_n} \nonumber\\
                 & = & \Delta t^P \frac{n - H}{(n - 1)H} \nonumber\\
                 & = & \Delta t_{min}\frac{1}{H}. \label{TMAX}
\end{eqnarray}
Figure \ref{theory} shows the ``optimal'' curve (using dynamic load balancing):
\begin{eqnarray}
  \frac{\Delta t^P}{\Delta t_{min}} = \frac{n-1}{n-H} \label{OPTCURVE}
\end{eqnarray}
and the one obtained without any load balancing:
\begin{eqnarray}
  \frac{\Delta t^P}{\Delta t_{max}} = \frac{n-1}{n-H}H \label{NOOPTCURVE}
\end{eqnarray}
for $n=4$. These curves will be compared with our experimental results.

\section{The Algorithm}

In this section we describe our algorithm to perform the dynamic load balancing,
based on the performance model described above. \\

\subsection{The Input:}

\begin{itemize}
  \item[-] A characteristic
  scale $L$ of the lattice (e.g.\ a side length of the lattice).
  \item[-] The number $n$ of processors of the multiprocessor system.
  \item[-] The total number
  of iterations $n_{iter}$ to be done by the simulation.
  \item[-] The number of iterations $n_{resize}$
  after which a resizing of the sublattices
           may take place.
  \item[-] A control parameter $\epsilon$ with $0 < \epsilon < 1$ to determine
           if a resizing should be done. \\
\end{itemize}

\subsection{The Output:}

\begin{itemize}
  \item[-] A dynamic resizing of the domains associated with each processor of
           the multiprocessor system, taking into account the actual resources
           of the processors. \\
\end{itemize}

\subsection{Formal Steps:}

\begin{itemize}
  \item[1.] Read the input.
  \item[2.] Introduce characteristic scales $L_i^{t_{MC}}$ of the sublattices
            with $i = 1,..,n$ and $t_{MC} = 1,..,n_{iter}/n_{resize}$,
            where $i$ denotes the processors and $t_{MC}$ counts the number of
            resizings that have been done.
  \item[3.] Calculate the initial characteristic sizes of the sublattices
            $L_i^{t_{MC}=0}$ for all processors according to
            $L_i^{t_{MC}=0} = \frac{L}{n}$.
  \item[4.] Associate each of the sublattices with one of the processors.
  \item[5.] Do on each processor $i = 1,..,n$ (in parallel)
  \begin{itemize}
    \item[\{]
    \item[-] Set $t_{MC} = 0$.
    \item[-] For $m = 1,.., n_{iter}$:
    \begin{itemize}
      \item[\{]
      \item[-] Perform iteration of the Monte Carlo update algorithm on the
               sublattice\footnote{Possibly including communication
               with other processors.}.
      \item[-] If $(m\  mod\ n_{resize}) = 0$ then
      \begin{itemize}
        \item[\{]
        \item[-] Measure the wall--clock time $\Delta t_{i}^{t_{MC}}$ spent on
                 processor $i$ for
                 doing the calculations
					  {\em excluding} the time spent for communications.
        \item[-] Calculate
                 \begin{equation}
                 P_{i}^{t_{MC}} := \frac{L_{i}^{t_{MC}}}
                             {\Delta t_{i}^{t_{MC}}}\label{ResourcesDet}
                 \end{equation}
                 to measure the actual resources of each node of the
                 multiprocessor system.
        \item[-] Communicate the results
		  to all processors of the multiprocessor system.
        \item[-] Calculate the new
		  characteristic sizes $L_{i}^{t_{MC}+1}$ of the
                 sublattices with
                 \begin{equation}
                 L_{i}^{t_{MC}+1} = \frac{P_{i}^{t_{MC}}}
                                         {\sum_{n=1}^{n} P_{n}^{t_{MC}}} L
                                    \label{ResizeOne}\ \ \ \
                 (i = 1,..,n-1)
                 \end{equation}
                 and
                 \begin{equation}
                 L_{n}^{t_{MC}+1} = L - {\sum_{n=1}^{n-1} L_{n}^{t_{MC}+1}}
                                           \label{ResizeTwo}.
                 \end{equation}
        \item[-] Resize the sublattices if
                 \begin{equation}
                 |L_{i}^{t_{MC}+1} - L_{i}^{t_{MC}}| > \epsilon
					  L \label{ResizeCond}.
                 \end{equation}

                 This step may include the communication of parts of the
                 sublattices between the processors and is certainly the
                 critical part of the algorithm. We will introduce an
                 algorithm for this resizing for a special case below.
        \item[-] Set $t_{MC} = t_{MC} + 1$.
        \item[\}]
      \end{itemize}
      \item[\}]
    \end{itemize}
    \item[\}]
  \end{itemize}
\end{itemize}

In the sequel we present an algorithm for
resizing the sublattices for the special
case that the splitting of the sublattices
takes place only in one dimension.
We use the {\em host--node}
(respectively {\em client--server}) parallel programming
paradigm (see \cite{Alt1}), associating each
sublattice with a server process and
leaving the more administration oriented tasks
(like reading the global parameters
of the simulation, starting the server processes, etc.) to the host
process\footnote{Of course these
tasks could also be done by the server processes,
resulting in the {\em hostless} paradigm of parallel programming. Therefore,
our algorithm could also be implemented in a {\em hostless} model
and our limitation to the {\em host--node} model is not a loss of generality.}.
Let us assume the size of the lattice in the direction of the splitting
to be $L$ and that the host process
holds arrays  $a[i]$ with ($i = 1,\ldots,n+1$)
for the ``Monte Carlo times'' $t_{MC}$ and $t_{MC} - 1$ containing
the first coordinate in that direction of the ``slice'' of the lattice
associated with each processor:
\begin{equation}
  a[1] = 1 \leq a[2] \leq ... \leq a[n+1] = L+1. \label{ArrayDef}
\end{equation}
(In terms of the constants $\{c_i\}$ this would mean
$ c_i = \frac{a[i+1]-a[i]}{L}$.)
Now the host process sends messages
containing instructions to the node processes
in two passes:
\begin{itemize}
  \item[1.] For $i = 2,\ldots,n$
  \begin{itemize}
    \item[\{]
    \item[] if ($(d := a_{t_{MC}}[i] - a_{t_{MC}-1}[i]) > 0$)
    \begin{itemize}
      \item[\{]
      \item[] send message to server $i$ telling it to
        send its ``first'' $d$ slices to processor $i-1$
      \item[\}]
    \end{itemize}
    \item[\}]
  \end{itemize}
  \item[2.] For $i = n-1,n-2,\ldots,1$
  \begin{itemize}
    \item[\{]
    \item[] if $(d := a_{t_{MC}}[i+1] - a_{t_{MC}-1}[i+1]) < 0$
    \begin{itemize}
      \item[\{]
      \item[] send message to server $i$ telling it to
         send its ``last'' $d$ slices to processor $i+1$
      \item[\}]
    \end{itemize}
    \item[\}]
  \end{itemize}
\end{itemize}

The node processes wait for messages from either the host process
or from neighbouring node processes. If there are not enough
slices available on a node process to be sent, the node process waits
for a message from a neighbour node process to receive additional slices.
The two pass algorithm prevents deadlocks.\\

If the resources of the processors of the multiprocessor system change
very rapidly, a multiple communication of data may be necessary and will
drastically reduce the efficiency of this algorithm. But this is consistent
with the fact, that our complete approach to dynamic load balancing is
anyhow only
valid for systems with moderatly varying resources, as was already pointed out
at the beginning of section 2, see (\ref{PSIMP}). \\

\section{Results for the Two--Dimensional Ising Model}

The above described algorithm has been implemented for the parallelized
simulation of the two--dimensional Ising model on a cluster of four
IBM RISC System/6000 -- 550 workstations \cite{Alt1}, using the PVM programming
environment \cite{PVM1, PVM2}. Here we have a two--dimensional
lattice which is divided into stripes.
The objects defined on the lattice sites are spins (i.e. binary variables)
and an iteration
defined on these objects consists e.g. of a Metropolis algorithm to generate
a new spin configuration on the lattice.
Each stripe is associated with one workstation. The characteristic scales of
the stripes are their widths and the characteristic scale of the lattice is
the sum of all widths. \\

\begin{figure}[h]
  \centering
  \leavevmode
  \epsffile{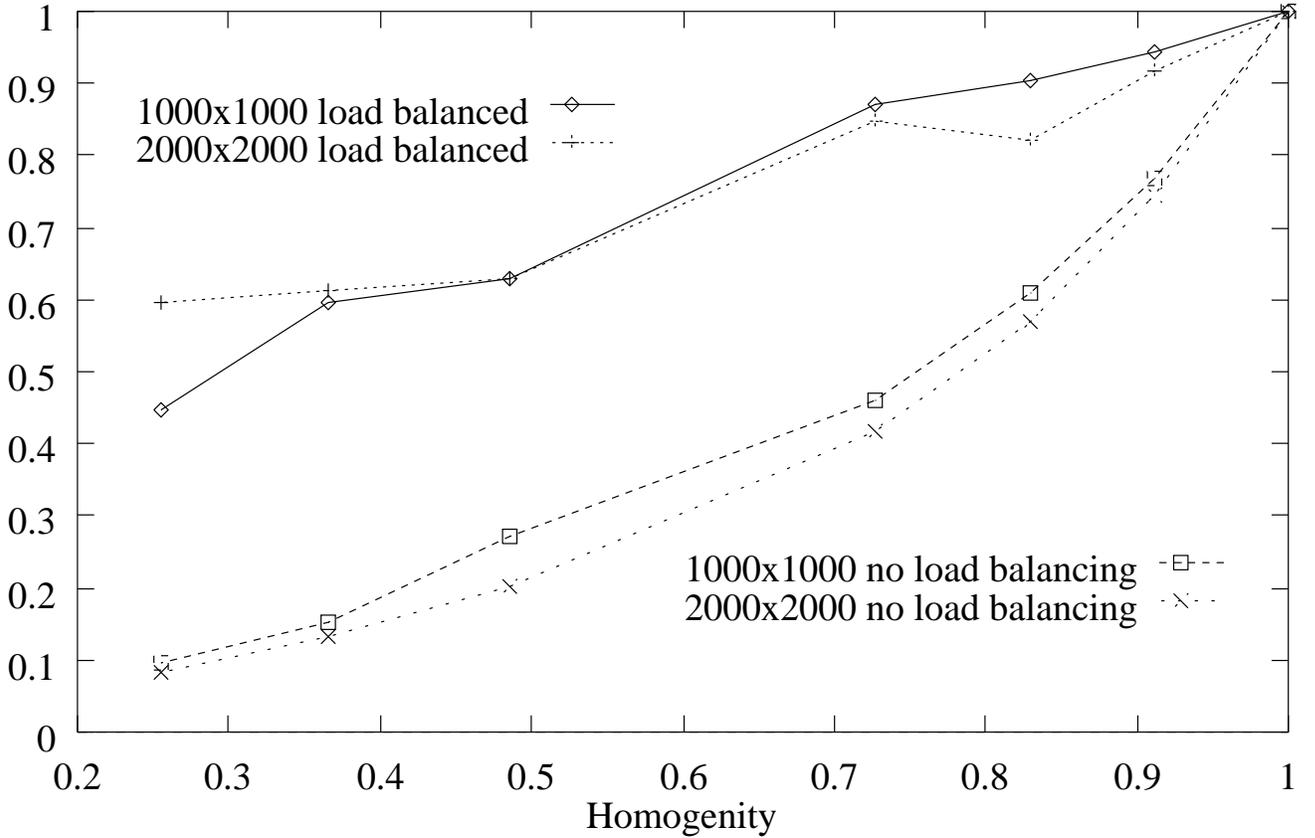}
  \caption{Performance measured for 4 processors with
           and without load balancing. \label{measurements}}
\end{figure}

The cluster being completely homogeneous, the heterogeneous situation
has been simulated by starting independent processes on one
or several nodes of the cluster. This allows the heterogenity of the
multiprocessor system
to be introduced in a controlled manner\footnote{During the measurements cited
below, the cluster has been dedicated to our application.}, i.e. to vary the
homogenity $H$ and
measure (\ref{OPTCURVE}) resp. (\ref{NOOPTCURVE}) as functions
of $H$. Our results are presented in figure \ref{measurements} for
a $1000 \times 1000$ and
a $2000 \times 2000$ lattice. One clearly sees the qualitative agreement with
the prediction of our performance model, see figure
\ref{theory}\footnote{Considering the fact, that we have not included the time
spent for communication in our model, a {\em quantitative} agreement between
the theoretical and measured performance cannot be expected. An inclusion
of the communication in our model would be very difficult and highly system
dependend, e.g. because system parameters like latency and bandwidth may
be be complicated functions of the homogenity $H$.}.
\begin{figure}[h]
  \centering
  \leavevmode
  \epsffile{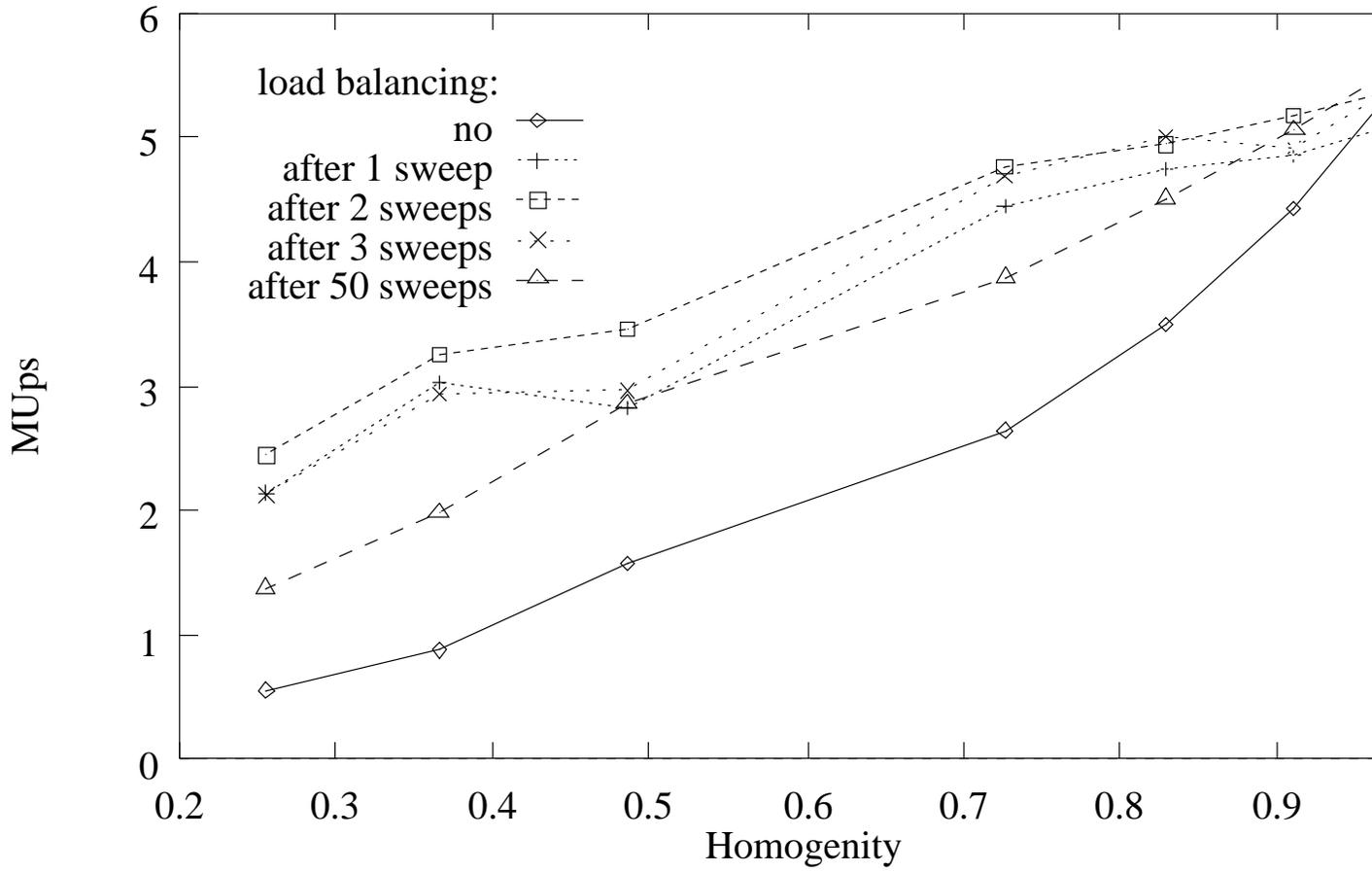}
  \caption{Performance measured in
  MUps for 4 processors for a $1000 \times 1000$
           lattice with
           and without load balancing being done after a certain number of
           sweeps. \label{mups1000}}
\end{figure}
\begin{figure}[h]
  \centering
  \leavevmode
  \epsffile{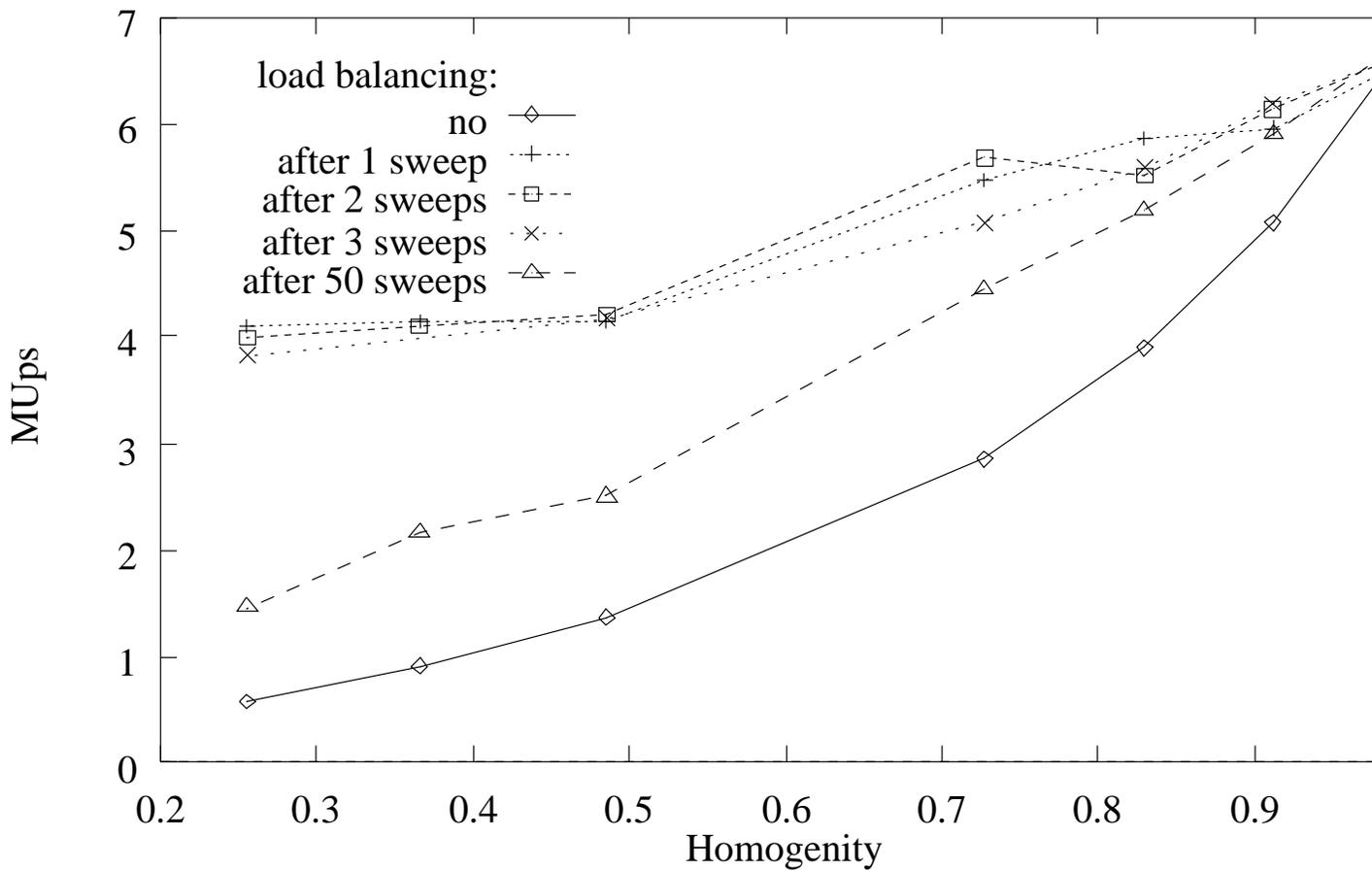}
  \caption{Performance measured in
  MUps for 4 processors for a $2000 \times 2000$
           lattice with
           and without load balancing being done after a certain number of
           sweeps. \label{mups2000}}
\end{figure}

A different point of view consists of looking at the (mega) updates done by the
Metropolis algorithm on each spin per second (``MUps'')\footnote{These ``MUps''
constitute a benchmark for spin models.}. These are presented for a
$1000 \times 1000$ and a $2000 \times 2000$ lattice as a function
of $H$ in figures \ref{mups1000} and \ref{mups2000} with the dynamic
load balancing being done after a certain
number of sweeps. It turns out, that the optimal number of sweeps between
the load balancing to be performed depends on the size of the problem. \\

\clearpage

\section{Summary}

We have introduced an algorithm for dynamic load balancing for synchronous
Monte Carlo simulations on a
heterogenous multiprocessor system with distributed
memory. Implementing this algorithm for the two--dimensional Ising model, we
have shown, that it may result
in a speedup of a factor 5 - 6 for the above described
class of geometrically parallelized algorithms.
In many cases, the implementation of the
algorithm is straight forward with only little overhead in calculation
and communication.
For homogeneous systems, almost no performance is lost because
the algorithm detects that no resizing is
necessary by applying (\ref{ResizeCond}).
For systems with slowly changing heterogenity\footnote{compared
to the time needed for one iteration (sweep)},
the algorithm converges very fast and
the requirements of the algorithm concerning the computing environment
are minimal: the system only has to provide a routine
to measure the wall--clock time; such a routine should be available on all
operating systems. \\

Considering the generality of the algorithm introduced above, it may also be
useful applied to problems other than Monte Carlo simulations,
e.g. in parallel iterative methods for solving linear or nonlinear equations
appearing in engineering
problems\footnote{Here the domain consists of a lattice,
with a matrix element being associated with each of the nodes of the lattice.}.

\end{document}